\documentclass[aps,prd,amsmath,amssymb,superscriptaddress,altaffillsymbol, preprintnumbers,nofootinbib,twocolumn]{revtex4}
\pdfoutput=1
\usepackage{amsthm}
\usepackage{graphicx}
\usepackage{subfigure}
\usepackage{tikz}
\usepackage{color}
\newcommand{\beq}{\begin{eqnarray}}
\newcommand{\eeq}{\end{eqnarray}}

\newcommand{\bmp}{\noindent\begin{minipage}{16cm}}
\newcommand{\emp}{\end{minipage}\vskip 7mm} 

\usepackage{dcolumn}
\usepackage{bm}
\usepackage{bbm}
\usepackage{subfigure}
\usepackage{slashed} 

\usepackage{lipsum}

\theoremstyle{definition}

\theoremstyle{plain}

\usepackage{epsfig}

\usepackage{hyperref}
\definecolor{rossoCP3}{cmyk}{0,.88,.77,.40}
\definecolor{verdeCP3}{rgb}{0.09765625, 0.57421875, 0.1015625}
\definecolor{bluCP3}{rgb}{0, 0.23, 0.67}
\hypersetup{colorlinks, bookmarksopen, bookmarksnumbered,citecolor=bluCP3, linkcolor=bluCP3, pdfstartview=FitH, urlcolor=bluCP3}
\def\lsim{\mathrel{\rlap{\lower4pt\hbox{\hskip1pt$\sim$}}
    \raise1pt\hbox{$<$}}}                
\def\gsim{\mathrel{\rlap{\lower4pt\hbox{\hskip1pt$\sim$}}
    \raise1pt\hbox{$>$}}}                

\newcommand{\bea}{\begin{eqnarray}}
\newcommand{\eea}{\end{eqnarray}}

\newcommand{\ba}{\begin{eqnarray}}
\newcommand{\ea}{\end{eqnarray}}

\newcommand{\be}{\begin{eqnarray}}
\newcommand{\ee}{\end{eqnarray}}

\begin{document}
\title{Earth's Stopping Effect in Directional Dark Matter Detectors}
%
\author{Chris Kouvaris}
\email{kouvaris@cp3.dias.sdu.dk} 
\affiliation{{\color{black} CP$^{3}$-Origins} \& Danish Institute for Advanced Study {\color{black} DIAS}, University of Southern Denmark, Campusvej 55, DK-5230 Odense M, Denmark}


\begin{abstract}

We explore the stopping effect that results from interactions between dark matter and nuclei as
the dark matter particles travel undergound towards the detector. Although this effect is negligible
for heavy dark matter particles, there is parameter phase space where the underground interactions of the dark matter particles with the nuclei can create observable differences in the spectrum. Dark matter particles that arrive on the detector from below can have less energy from the ones arriving from above. These differences can be potentially detectable by upcoming directional detectors. This can unveil a large amount of information regarding the type and strength of interactions between nuclei and light dark matter candidates.

\preprint{CP3-Origins-2015-038 DNRF90, DIAS-2015-38.}
 \end{abstract}

\maketitle
\section{Introduction}

There is strong evidence for the existence of Dark Mat-
ter (DM) nowdays. Searches for DM include efforts for
laboratory production (e.g. LHC), possible indirect sig-
nals from the galaxy and beyond (e.g. due to annihilation
or decay of DM to conventional photons or other Standard Model paticles), and direct detection where underground detectors could potentially register rare collisions
between an incoming DM particle and a nucleus in the
detector. Current direct search experiments can register
events with a particular recoil energy, but they cannot
identify the direction of the recoil. However, a new generation of experiments that can detect also the direction
of the recoil is on the way~\cite{Battat:2014van,Daw:2011wq,Riffard:2013psa,Santos:2013hpa,Battat:2013gma,Monroe:2012qma,Miuchi:2010hn,Vahsen:2011qx}. The directional detection
of these experiments is based on time projection diffuse gas chambers that have the capability of recostructing
the nuclear recoil track, giving thus information about
the direction of the incoming DM particle. Additionally
there are directional detectors that are based on different techniques such as nuclear emulsion on solid detectors~\cite{Naka:2011sf}, DNA detectors~\cite{Drukier:2012hj}, and DM-electron scattering
in crystals~\cite{Essig:2011nj}. Although the above experiments are not
yet competitive in setting DM limits with respect to the
current non-directional underground detectors, they will
soon be able to start probing interesting DM parameter
space.

Directional DM detectors once competitive to non-directional ones, can provide an immense amount of information that cannot be obtained otherwise. 
Conventional detectors register counts with particular recoil energies. Although the number of expected counts depends
on the velocity distribution of the DM halo particles, it is
hard to extract useful information regarding the form of
the distribution function due to the fact that for a particular amount of nuclear recoil energy, all DM particles with velocities above a specific value could produce the
observed recoil. The number of counts in the detector is
proportional to an integral of the DM velocity distribution, thus making hard to reveal the exact shape of this
distribution. On the contrary in directional detectors,
the directional rate of counts per recoil energy is associated with the Radon transformation of the DM velocity
distribution~\cite{Gondolo:2002np}. This can be in principle inverted and
therefore one can obtain direct correspondence between
the velocity distribution and the amount of registered
counts on the detector. Furthermore, directional detection can help on two other fronts. On the one hand, it is
much easier to eliminate background contamination with
directional detectors. Known sources of contamination
such as for example the sun can be easily eliminated.
On the other hand, directional detectors can reveal information about possible substructure of the DM halo.
Directional detectors could shed light on the possibility
of DM streams and dark discs in the galaxy. This would
be something almost impossible to probe with conventional detectors.

The recoil energy spectrum of DM scattering off nuclei in directional detectors has been studied extensively
first in~\cite{Gondolo:2002np} and later in~\cite{Morgan:2004ys,Green:2006cb,Alenazi:2007sy,Green:2007at,Green:2010zm,Billard:2011zj,Grothaus:2014hja,Kavanagh:2015aqa,Laha:2015yoa,Bozorgnia:2012eg,Bozorgnia:2011vc}. In all the above papers,
the energy recoil spectrum has been studied for the two
generic cases of spin-independent and spin-dependent
DM-nucleon contact interactions. An extension to more
generic non-relativistic scattering operators was studied
in~\cite{Catena:2015vpa}.

In this paper we study the stopping effect of the earth in directional DM detectors. DM particles can arrive at the detector from different angles, having traveled different
 distances underground. Although DM particles are expected to interact feebly with nucleons, as it was pointed out in~\cite{Kouvaris:2014lpa}, there is DM parameter space especially for light DM candidates where DM-nuclei interactions as the particle travels underground might have an observable effect on the recoil energy spectrum of the detectors. There are two ways that underground DM-nuclei interactions can affect the spectrum. The first one is that particles traveling larger distances underground,
might lose enough energy due to interactions, so by the
time they reach the detector might not have enough energy to create a recoil above the threshold of the detector.
This clearly creates an asymmetry between the amount of registered counts in the detector coming from above
and from below. Additionally, DM-nuclei interactions
can cause also the opposite effect for heavy enough DM
particles. DM particles that interact undeground as they
travel towards the detector, slow down. This reduction in
the velocity might increase the DM-nucleus cross section
and therefore the probability of detecting it. This is because in a variety of models the cross section is inversely
proportional to some power of the velocity (e.g. for contact interactions  $\sigma \sim 1/v^2$). The study of this stopping
effect of the earth was studied in the context of conventional non-directional detectors in~\cite{Kouvaris:2014lpa}. In this case, since
there is no way to know the direction of the recoiled nucleus, the effect can be seen indirectly via the observation
of a diurnal modulated signal. As it was demonstrated
in~\cite{Kouvaris:2014lpa}, since the earth moves with a nonzero velocity with
respect to the rest frame of the galaxy, a daily varying
DM signal is created because as the earth rotates around
its own axis, the DM particles coming from the direction
of the DM wind travel different distances underground
at different times during a sidereal day. The observation of such a diurnal modulated signal can reveal information about the nature of DM-nucleon interactions.
Additionally for detectors placed on shallow sites, this
technique might be one of the few options available to
probe light DM parameter space with long range forces
that is currently inaccessible to detectors. Diurnal modulation has been investigated in the past in the context of
Strongly Interacting Massive Particles~\cite{Zaharijas:2004jv,Collar:1992qc} and mirror DM~\cite{Foot:2011fh,Foot:2014osa}, as well as experimentally in the DAMA
Collaboration~\cite{Bernabei:2014jnz,Bernabei:2015nia}.

 In this paper we study the same
stopping effect of the earth in a more direct way, which
is in the context of directional DM detectors. One should not have to
rely on a diurnal modulated signal in order to probe the
asymmetry in the spectrum between DM particles scattering from below and above. The paper is organised as
follows: in section II we review the stopping power of
DM particles due to DM-nuclei interactions. In section
III we will derive the formalism for the directional recoil
spectrum and we will present our results.

\section{Nuclear Stopping}
 DM particles can lose energy by interacting with nuclei or electrons as they travel underground towards the
detector. DM particles from the halo do not have sufficient energy to ionize atoms as they travel underground.
They can lose energy either by interactions with nuclei,
or if allowed, by interactions with electrons. The
latter can be either in the form of DM interactions with
electrons in metallic layers of the earth, or in the form
of DM-electron interactions that result in atomic excitations~\cite{Kouvaris:2014lpa}. The determination of the most effective mode of decelerating DM particles depends strongly on the type of DM-nucleus interactions as well as the precise geological composition of the earth. For example contact
or long range forces between DM and nuclei can result to different degrees of DM deceleration inside the earth.
 In this work here, we are going to consider only nuclear
stopping. This is because nuclear stopping is quite insensitive to the geological composition of the earth. For example DM-electron interactions in metallic layers of the
earth can give significant amounts of stopping because
electrons there behave as a free Fermi gas that does not
have an energy gap and therefore it can subtract energy from incoming DM particles by small bits at the
time. However, they are model dependent, depending
strongly on the geological morphology of the earth. For
simplicity, we are going to consider contact spin-independent DM-nucleon interactions here. Our goal is to make a first generic estimate on the possibility of observing the stopping effect of the earth in directional detectors. Additional stopping modes for DM particles can
only enhance the effect. Moreover we are going to assume
a flat density for the earth of $\rho_e=5.5\text{gr}/\text{cm}^3$. This will
enable us to obtain more transparent results regarding
the spectrum of the recoiled energy in directional detectors.

For a DM particle moving through a medium, the energy loss per distance traveled is given by
\be
\frac{dE}{dx}=-\sum_i n_{N_i} \int_{E_R^{\text{min}}}^{E_R^{\text{max}}} \frac{d\sigma_i}{dE_R}E_R dE_R, \label{stop}
\ee
where $n_{N_i}$ is the number density of nuclei $N_i$ and $d\sigma_i/dE_R$ is the differential cross section between $N_i$ and DM which in the case of contact spin-independent interactions is given by 
\be
\frac{d \sigma}{dE_R}=\frac{m_N \sigma_N}{2 \mu_N^2 v^2}F^2(E_R)=\frac{m_N \sigma_p A^2}{2 \mu_p^2 v^2}F^2(E_R), \label{cross}
\ee
where $m_N$ is the mass of the target nucleus, $A$ the number of nucleons in the nucleus and $\mu_p$ ($\mu_N$) the reduced mass between DM and proton (nucleus). $\sigma_N$ and $\sigma_p$
are correspondingly the DM-nucleus and DM-nucleon cross sections. $F^2(E_R)$ is the usual form factor that accounts for loss of coherence. We choose a simple form factor of the form
\be
F^2(E_R)=e^{-E_R/Q_0},
\ee
where $Q_0=3/(2m_N r_0^2)$ and $r_0=0.3+0.91(m_N)^{1/3}$ is the radius of the nucleus measured in femtometers when $m_N$ is in GeV~\cite{Copi:2000tv}.
The sum in Eq.~(\ref{stop}) runs over all
the elements found in the earth. However, as it was argued in~\cite{Kouvaris:2014lpa}, oxygen is the element that gives the highest
contribution at least for contact spin-independent interactions (being abundant by   48$\%$ in the earth) and
therefore this will be the single element we are going
to consider in the following calculation. This simplifies
the formulas significantly while introducing only small
errors in the estimate of the stopping. Once again,
the error is in the right direction, i.e. extra contributions from other elements can only enhance the stopping effect of the earth we study here. The integral
of Eq.~(\ref{stop}) has lower and upper limits $E_R^{\text{min}}$ and $E_R^{\text{max}}$ respectively.  $E_R^{\text{max}}=4m_X m_N E/(m_X+m_N)^2$ is the maximum recoil energy given a DM particle of energy $E$ ($m_X$ being the DM mass). For  perfect contact interaction $E_R^{\text{min}}=0$. However in a realistic case, contact
interactions might result by integrating out heavy mediators. For example in a Yukawa type of interaction
between DM and nucleons where a mediator of mass $m_{\phi}$ 
 is exchanged, DM and nucleon should come closer than a distance  $m_{\phi}^{-1}$. This requires the exchange of a mediator with energy determined by uncertainty principle of at least  $E_R^{\text{min}}=m_{\phi}^2/(2 m_N)$. Upon writing $v^2=2E/m_X$, Eq.~(\ref{stop}) can be integrated to
\be
\frac{dE}{dx}=-\frac{2 n_N \sigma_p A^2 \mu_N^4 E}{m_X m_N \mu_p^2},
\ee
where we used $E_R^{\text{max}}=4m_X m_N E/(m_X+m_N)^2$,  $E_R^{\text{min}}<<E_R^{\text{max}}$ and $E_R^{\text{max}}<<Q_0$ (the last is especially true for low DM masses that we are particularly interested).  The final trivial integration upon the assumption
that the density and composition of the earth is constant, gives
\be
\ln\frac{E_{in}}{E_f}=\frac{2 n_{N_s} \sigma_p A_s^2 \mu_{N_s}^4 L}{m_X m_{N_s}\mu_p^2}, \label{ln}
\ee
where $E_{in}$ and $E_f$  are respectively the initial and final
kinetic energies of the DM particle and $L$  is the total
length traveled underground. Note that we have added
an index $s$  in $A$ , $n_N$ , $m_N$  and $\mu_N$  in order to distinguish
the nucleus responsible for the deceleration of the DM
particles (i.e. oxygen)  from the nucleus that
serves as a target in the detector (that can be an element
different from oxygen). Eq.~(\ref{ln}) can be rewritten in terms
of velocities as
\be
v'=v e^{-\Delta L}, \label{vv1}
\ee
where $v'$ and $v$ are the final velocity (after the particle has traveled $L$ underground) and initial velocity (before the particle enters the earth) of the DM particle. $\Delta$ is
\be
\Delta=\frac{n_{N_s} \sigma_p A_s^2 \mu_{N_s}^4}{m_X m_{N_s} \mu_p^2}.
\ee
\section{Recoil Energy Spectrum}
We are going to consider now the  energy recoil spectrum in directional detectors. Generally, the rate of counts (counts per time) per recoil energy per solid angle is~\cite{Gondolo:2002np}
\be
 \frac{d^{2}R}{dE_{R}d\Omega_{q}}=N_{T}n_{\chi}\int\frac{d^{2}\sigma}{dE_{R}d\Omega_{q}}f(v)vd^{3}v.   \label{spec}
\ee
$E_{R}$ is the recoil energy, $\Omega_{q}$ a solid angle around the direction of $\hat{q}, f(v)$ is the DM velocity distribution, $N_{T}$ is the number of nuclei targets in the detector and
$n_{\chi}=0.3\text{GeV cm}^{-3}/m_X$ is the DM number density in the earth. The directional differential cross section is related to the non-directional one as
\be
 \frac{d^{2}\sigma}{dE_{R}d\Omega_{q}}=\frac{1}{2\pi}\delta(\cos\theta-\frac{v_{\min}}{v'})\frac{d\sigma}{dE_{R}},   \label{dircross}
\ee
where $v_{\min}=\sqrt{m_{N}E_{R}/(2\mu_{N}^{2})}$ is the minimum velocity that can produce a recoil energy $E_{R}$. $\mu_{N}$ here is the reduced mass between DM and the target nucleus of the detector $N$, and $\theta$ is the angle between the velocity of the DM particle and the direction of the recoiled nucleus $\hat{q}$. Using Eqs.~(\ref{cross}) and (\ref{dircross}) in (\ref{spec}), we get
\be
 \frac{d^{2}R}{dE_{R}d\Omega_{q}}=\kappa\int\frac{1}{v'^{2}}\delta(\cos\theta-\frac{v_{\min}}{v'})f'(x',v')v'd^{3}v', \label{dirrate1}
\ee
where $\kappa=N_{T}n_{\chi}m_{N}\sigma_{p}A^{2}F^{2}(E_{R})/(4\pi\mu_{p}^{2})$. One should keep in mind that $N$ and $A$ in the above equation refer to the target element of the detector. The reader should also notice that the $1/v^{\prime 2}$ dependence inside the integral comes from the fact that the scattering between DM and nucleus takes place with a DM velocity $v'$ which is smaller than the velocity of DM before enters the earth and is given by Eq.~(\ref{vv1}). Similarly the flux is given by the distribution of DM $f'(x',v')v'$ at the location of the detector, which is not the same as the DM flux at the surface of the earth $f(v)v$ (where no DM deceleration has taken place). We can find a relation between $f'(x',v')$ and $f(v)$ by using Liouville theorem. Let us approximately consider that DM moves on a straight line underground and DM-nuclei interactions inside the earth decelerate the particle but they don't deflect it from its path. The distribution of DM as it enters the earth is governed by the Boltzmann equation
\be
\frac{\partial f}{\partial t}+v_{i}\frac{\partial f}{\partial x_{i}}+a_{i}\frac{\partial f}{\partial v_{i}}=0,  
\ee
where we assumed that no collisions take place among DM particles. The acceleration $a_{i}$ results from the force induced by the DM-nuclei undeground scatterings and this force is treated as an external one. Since we are in- terested in steady state solutions, one can set $\partial f/\partial t=0.$ This means that $f(x_{i}(t),v_{i}(t))$ remains constant along the trajectory of a DM particle (which is a straight line underground). This is a manifestation of the Liouville theorem and therefore $f(v)=f'(x',v')$, i.e. the distribu- tion at the detector is equal to the one before DM enters the earth. Using this fact as well as $d^{3}v'=e^{-3\Delta L}d^{3}v$ (see Eq.~(\ref{vv1})) we can rewrite Eq.~(\ref{dirrate1}) as
\be
 \frac{d^{2}R}{dE_{R}d\Omega_{q}}=\kappa\int\delta(\cos\theta-\frac{v_{\min}}{v}e^{\Delta L})\frac{f(v)}{v}e^{-2\Delta L}d^{3}v. \label{rateX}
\ee
This is the main formula we are going to use in order to probe the stopping effect of the earth. In particular, we are going to consider the asymmetry in the directional rate between the two directions that give the largest possible difference, i.e. $\hat{q}=\hat{n}$ and $\hat{q}=-\hat{n}$, where $\hat{n}$ is the direction from the center of the earth to the position of the detector. This two directions correspond to particles that travel the shortest distance underground ($\hat{q}=-\hat{n})$ and the largest one $(\hat{q}=\hat{n})$.

We are going to use a truncated Maxwell distribution
\be
f(v)= \frac{1}{\mathcal{N}}\exp \left [-\frac{(\vec{v}+\vec{v}_{e})^{2}}{v_{0}^{2}} \right ],~v<v_{\mathrm{e}\mathrm{s}\mathrm{c}}+v_{e}, \label{trancMax}
\ee
where $\mathcal{N}$ is a normalization constant, $v_{e}$ is the velocity of the earth with respect to the rest frame of the DM halo, and $ v_{\mathrm{e}\mathrm{s}\mathrm{c}}=550\mathrm{k}\mathrm{m}/\sec$ is the escape velocity from Milky Way. It is understood that the velocity distribution is in the labaratory frame (boosted by $\vec{v}_{e}$). The length trav- eled underground by a DM particle is given by
\bea
 L&=&(R_{\oplus}-\ell_{D})\cos\psi   \nonumber \\
&+&\sqrt{(R_{\oplus}-\ell_{D})^{2}\cos^{2}\psi-(\ell_{D}^{2}-2R_{\oplus}\ell_{D})}, \label{length}
\ee
where $\cos\psi=\hat{v}\cdot \hat{n}$ represents the angle between the DM velocity and the upper direction of the detector $\hat{n}$.  $R_\oplus$ and $\ell_{D}$ are the earth's radius and the depth of the detector respectively.

Let us consider in some detail the different directions and angles involved in the problem. Following~\cite{Kouvaris:2014lpa} we define $\theta_{l}$ to be the latitude of the detector, and we choose the $z$-axis with direction south-north pole. $\alpha$ is the angle between $\vec{v}_{e}$ and the $z$-axis. We choose the orientation of the $x-y$ plane so $\vec{v}_{e}$ lies along the $x-z$ plane. In this reference system choice we have the following relations
\be
\hat{n}=\hat{x}\cos\theta_{l}\cos\omega t+\hat{y}\cos\theta_{L}\sin\omega t\pm\hat{z}\sin\theta_{l},   \label{nn}
\ee
\be
\hat{v}_{e}=\hat{x}\sin\alpha+\hat{z}\cos\alpha,   
\ee
where the $\pm$ corresponds to the north and south hemisphere. We have chosen $t=0$ the time where $\vec{v}_{e}$ and $\hat{n}$
align as much as possible, i.e. $\hat{n}$ is along the $x-z$ plane. Eq.~(\ref{trancMax}) can now be rewritten as
\be
f(v)=\frac{1}{\mathcal{N}}e^{-\frac{v^2+v_e^2}{v_0^2}}e^{-\frac{2v v_e}{v_0^2}\cos\delta},  
\ee
where $\delta$ is the angle between $\vec{v}$ and $\vec{v}_{e}$. In order to find $\delta$ we express the WIMP velocity $\vec{v}$ as
\be
\vec{v}=v(\hat{x}\sin\theta\cos\phi+\hat{y}\sin\theta\sin\phi+\hat{z}\cos\theta),  \label{vdef}
\ee
where we use the usual polar angles $\theta$ and $\phi$ to charac- terize $\vec{v}$. The angle $\delta$ now reads
\be
\cos \delta=\hat{v}\cdot\hat{v}_{e}=\sin\alpha\sin\theta\cos\phi+\cos\alpha\cos\theta. \label{delt}
\ee
We are interested in the difference on the directional detection rate between the direction $\hat{q}=\hat{n}$, i.e. the direc- tion coming from the center of the earth to the detector and $\hat{q}=-\hat{n}$ (the opposite one). Practically speaking, we probe the asymmetry in the directional detection rate be- tween events in the detector that come from below and from above. In the case where $\hat{q}=\hat{n}$, one can notice that $\psi=\theta$. Let us define $ y=\cos\theta$, and $y_+$ that satisfies
\be
y_+=\frac{v_{\min}}{v}e^{\Delta L_{+}(y_+)},  \label{y1}
\ee
where $L_{+}(y)=L$ defined in Eq.~(\ref{length}) (with $\cos \theta\rightarrow y$). $y_+$ in Eq.~(\ref{y1}) is nothing else but the value of $y$ (or $\cos\theta$) that makes the arguement inside the delta function of Eq.~(\ref{rateX}) zero. Eq.~(\ref{rateX}) can be written as
\be
 \frac{d^{2}R}{dE_{R}d\Omega_{n}}=\frac{\kappa}{\mathcal{N}}\int e^{-\frac{v^{2}+v_{\mathrm{e}}^{2}+2vv_{\mathrm{e}}\cos\delta}{v_{0}^{2}}}ve^{-2\Delta L_{+}}dvd\phi. 
\ee
Recall that $\exp[-2\Delta L_{+}]=v_{\min}^{2}/(v^{2}y_{+}^{2})$ (from Eq.~(\ref{y1})) and that $y_+$ is a function of $v$ and $E_{R}$. Using this and Eq.~(\ref{delt}), we have
\begin{widetext}
\bea
 \frac{d^{2}R}{dE_{R}d\Omega_{n}}&=&\frac{\kappa}{\mathcal{N}}\int\exp \left [-\frac{v^{2}+v_{e}^{2}+2vv_{e}(y_+\cos\alpha+\sin\alpha\sqrt{1-y_{+}^{2}}\cos\phi)}{v_{0}^{2}} \right ]\frac{v_{\min}^{2}}{vy_{+}^{2}}dvd\phi \nonumber \\
&=& \frac{2\pi\kappa}{\mathcal{N}}\int_{v_{1}}^{v_{\mathrm{e}\mathrm{s}\mathrm{c}}+v_{\mathrm{e}}}\exp \left [-\frac{v^{2}+v_{e}^{2}+2vv_{e}y_+\cos\alpha}{v_{0}^{2}} \right ]\mathcal{I}_{0}\left (\frac{2vv_{e}}{v_{0}^{2}}\sin\alpha\sqrt{1-y_{+}^{2}}\right)\frac{v_{\min}^{2}}{vy_{+}^{2}}dv,   \label{difr}
\eea
\end{widetext}
where we have used $\sin\theta=\sqrt{1-y_{+}^{2}}$, and we have integrated over $\phi$ in the second line. $\mathcal{I}_{0}$ is the modified Bessel function of the first kind. The minimum velocity $v_{1}$ is the solution of $v_{1}=v_{\min}\exp\{\Delta L_{+}[y_{+}(v_{1},E_{R})]\}.$ As we mentioned $L_+$ is a function of $y_+$ which is a function of $v$. A comment is in order here. $\theta$ is the angle between the recoil direction $\hat{q}$ and $\hat{v}$. Since in this case $\hat{q}=\hat{n}, \theta$ is the angle between $\hat{n}$ and $\hat{v}$. However, as it can be seen from Eq.~(\ref{nn}), the vector $\hat{n}$ has $\hat{x}$ and $\hat{y}$ time varying components, while $\hat{v}$ in Eq.~(\ref{vdef}) is expressed in spherical coordinates around the $\mathrm{z}$-axis. Since we have
chosen $\hat{q}=\hat{n}$, $\cos\theta$ should express the angle between $\hat{n}$ and $\hat{v}$ and not the angle between $\hat{z}$ and $\hat{v}$. In order to simplify our calculation and without introducing a big error in our estimate, we take the time average value of $\hat{n}$ which concides with $\hat{z}$. In other words within our approximation we have assumed that $\hat{n}=\hat{z}$ and therefore $\theta$ of Eq.~(\ref{vdef}) coincides with the definition of $\theta$ being the angle between $\hat{v}$ and the recoil direction $\hat{q}$.

Let us look now on the directional rate from above ($\hat{q}=-\hat{n}$). In this case $\psi=\pi-\theta$ (and $\cos \psi=-\cos\theta$) and the distance traveled underground of Eq.~(\ref{length}) becomes
 \bea 
L_{-}&=&-(R_{\oplus}-\ell_{D})\cos\theta \nonumber \\
&+&\sqrt{(R_{\oplus}-\ell_{D})^{2}\cos^{2}\theta-(\ell_{D}^{2}-2R_{\oplus}\ell_{D})}.
\eea
Since $\hat{q}=-\hat{n}\simeq-\hat{z}$ (the last equality holding as a time average of Eq.~(\ref{nn})), one should express $\hat{v}$ in spherical coordinates but with $\hat{z}\rightarrow-\hat{z}$. In this case the angle $\delta$ between $\hat{v}$ and $\hat{v}_{e}$ picks up a relative minus sign in the second term of Eq.~(\ref{delt}), thus reading
\be
\cos \delta=\hat{v}\cdot\hat{v}_{e}=\sin\alpha\sin\theta\cos\phi-\cos\alpha\cos\theta.
\ee
The directional recoil rate can be written as
\begin{widetext}
\bea
\frac{d^{2}R}{dE_{R}d\Omega_{-n}}&=&\frac{\kappa}{\mathcal{N}}\int\exp\left [-\frac{v^{2}+v_{e}^{2}-2vv_{e}(y_-\cos\alpha-\sin\alpha\sqrt{1-y_{-}^{2}}\cos\phi)}{v_{0}^{2}}\right ]\frac{v_{\min}^{2}}{vy_{-}^{2}}dvd\phi \nonumber \\
&=& \frac{2\pi\kappa}{\mathcal{N}}\int_{v_{2}}^{v_{\mathrm{e}\mathrm{s}\mathrm{c}}+v_{\mathrm{e}}}\exp\left [-\frac{v^{2}+v_{e}^{2}-2vv_{e}y_{-}\cos\alpha}{v_{0}^{2}}\right ]\mathcal{I}_{0}\left (\frac{2vv_{e}}{v_{0}^{2}}\sin\alpha\sqrt{1-y_{-}^{2}}\right )\frac{v_{\min}^{2}}{vy_{-}^{2}}dv,  \label{difr2}
\eea
\end{widetext}
where $y_{-}$ is defined as the number that satisfies
\be
y_-=\displaystyle \frac{v_{\min}}{v}e^{\Delta L_{-}(y_-)}.
\ee
$v_{2}$ is defined as the solution of $v_{2}=v_{\min}\exp\{\Delta L_{-}[y_{-}(v_{2},E_{R})]\}$. In the second line of the equation we have performed the integration over $\phi$. We are interested in the asymmetry on the directional recoil rate between $\hat{n}$ and $-\hat{n}$. However the two direc- tional rates are not equal in the first place, even if we ignore the stopping effect completely. Since the earth is moving with respect to the rest frame of the DM halo, $\vec{v}_{e}$ defines a direction that breaks isotropy. The perspective of probing the forward-back asymmetry using directional detectors has been explored thorougly~\cite{Morgan:2004ys,Kavanagh:2015aqa,Copi:2000tv,Copi:1999pw,Cerdeno:2010jj}. As a first step, we would like to estimate how big is the rate asymmetry between the direction $\hat{n}$ and $-\hat{n}$ due to the stopping effect we study compared to the pure forward-backward asymmetry due to the DM wind. This will gives us a sense of how easily this effect can be probed in directional detectors in the near future. Let us now calculate the forward-backward asymmetry due to the motion of the earth inside the galaxy. Following the steps from Eq.~(\ref{rateX}) to (\ref{difr}) and upon ignoring the stopping effect $(\mathrm{i}.\mathrm{e}.\ \Delta=0)$ we can derive the forward- backward asymmetry (i.e. the asymmetry between the directions $\hat{v}_{e}$ and $-\hat{v}_{e}$) as
\begin{widetext}
\be
 \Delta_{0}=\frac{d^{2}R}{dE_{R}d\Omega_{-v_{\mathrm{e}}}}-\frac{d^{2}R}{dE_{R}d\Omega_{v_{\mathrm{e}}}}=\frac{4\pi\kappa}{\mathcal{N}}\int_{v_{\min}}^{v_{\mathrm{e}\mathrm{s}\mathrm{c}}+v_{\mathrm{e}}}\exp \left [-\frac{v^{2}+v_{e}^{2}}{v_{0}^{2}}\right]\sinh\left [\frac{2v_{e}v_{\min}}{v_{0}^{2}}\right ]vdv. \label{asym2}
\ee
\end{widetext}
We can now estimate the significance of the stopping effect with respect to the forward-backward asymmetry by considering the following ratio
\be
R=\displaystyle \frac{\frac{d^{2}R}{dE_{R}d\Omega_{-n}}-\frac{d^{2}R}{dE_{R}d\Omega_{n}}}{\Delta_{0}}.
\ee
There is also another meaningful comparison we can make. We can compare the asymmetry due to the stopping effect compared to the pure asymmetry created in the flux by the DM wind evaluated in the up and down directions of the detector. In other words we get an estimate of the relevant importance of the stopping effect compared to
that of the velocity by considering
\be
R'=\displaystyle \frac{\Delta R_{s}-\Delta R_{0}}{\Delta R_{s}},
\ee
where
\be
\Delta R_{s}=\frac{d^{2}R_{s}}{dE_{R}d\Omega_{-n}}-\frac{d^{2}R_{s}}{dE_{R}d\Omega_{n}}, 
\ee
\be
\Delta R_{0}=\frac{d^{2}R_{0}}{dE_{R}d\Omega_{-n}}-\frac{d^{2}R_{0}}{dE_{R}d\Omega_{n}}.
\ee
The indices ``$s$'' and ``$0$'' refer to the directional recoil rates with stopping and after having ignored the stopping effect of the undeground atoms respectively. The latter is given by Eqs.~(\ref{difr}) and (\ref{difr2}) once we set $\Delta=0$, $y_+= y_{-}=v_{\min}/v$, and $v_{1}=v_{2}=v_{\min}$
\begin{widetext}
\be
 \frac{d^{2}R_{0}}{dE_{R}d\Omega_{-n}}-\frac{d^{2}R_{0}}{dE_{R}d\Omega_{n}}=\frac{4\pi\kappa}{\mathcal{N}}\int_{v_{\min}}^{v_{\mathrm{e}\mathrm{s}\mathrm{c}}+v_{e}}\exp \left [-\frac{v^{2}+v_{e}^{2}}{v_{0}^{2}} \right ]\sinh \left [\frac{2v_{e}v_{\min}\cos\alpha}{v_{0}^{2}}\right ]\mathcal{I}_{0} \left (\frac{2vv_{e}}{v_{0}^{2}}\sin\alpha\sqrt{1-\frac{v_{\min}^{2}}{v^{2}}} \right)vdv. \label{asym1}
\ee
\end{widetext}
One can notice that by setting $\alpha=0$, Eq.~(\ref{asym1}) reduces to Eq.~(\ref{asym2}). In addition to the previous ratios, it is important to estimate how big is the asymmetry compared to the total recoil rate, i.e. the number of counts per recoil energy after we integrate over the whole $ 4\pi$ solid angle. This can be probed by the ratio
\be
 \Delta R=\frac{\Delta R_{s}-\Delta R_{0}}{dR_{0}/dE_{R}}\delta\Omega.  
\ee
It is understood that $dR_{0}/dE_{R}$ is the total rate that pro- duces recoil energy $E_{R}$ (upon ignoring the stopping effect), i.e. the total non-directional rate after one inte- grates over the whole solid angle of $4\pi$. $\delta\Omega$ is the solid angle resolution for a typical directional detector. We take it here to be the solid angle of a cone with angle opening of $\pi/6$, i.e. $\delta\Omega=2\pi(1-\cos[\pi/6])$. $dR_{0}/dE_{R}$ can be easily estimated
\begin{widetext}
\bea
\frac{dR_{0}}{dE_{R}}&=&\frac{2\pi\kappa}{\mathcal{N}}\int\exp \left[-\frac{v^{2}+v_{e}^{2}+2vv_{e}\cos\theta}{v_{0}^{2}}\right ] vdvd\cos \theta d\phi \nonumber \\
&=& \frac{\pi^{5/2}\kappa v_{0}^{3}}{\mathcal{N}v_{e}} \left (\text{erf} \left [\frac{v_{\mathrm{e}\mathrm{s}\mathrm{c}}}{v_{0}} \right] -\text{erf} \left [ \frac{2v_{e}+v_{\mathrm{e}\mathrm{s}\mathrm{c}}}{v_{0}}\right ]+ \text{erf} \left [ \frac{v_{e}-v_{\min}}{v_{0}}\right ]+ \text{erf} \left [\displaystyle \frac{v_{e}+v_{\min}}{v_{0}} \right]\right).
\eea
\end{widetext}
$\Delta R$ is an important parameter because it reflects the amount of data needed in order to probe the stopping effect of underground atoms on DM. It is the difference in the amount of events detected in a detector with a direction in the recoil within a cone (calibrated to a typical angle of $\pi/6$) pointing down and a cone pointing up, after  subtracting the amount of the asymmetry due solely to the velocity of the earth with respect to the rest frame of the DM halo, over the total number of events (from all directions).
\begin{figure}[h!]
\begin{center}
\includegraphics[width=.4
\textwidth, height=0.3 \textwidth
]{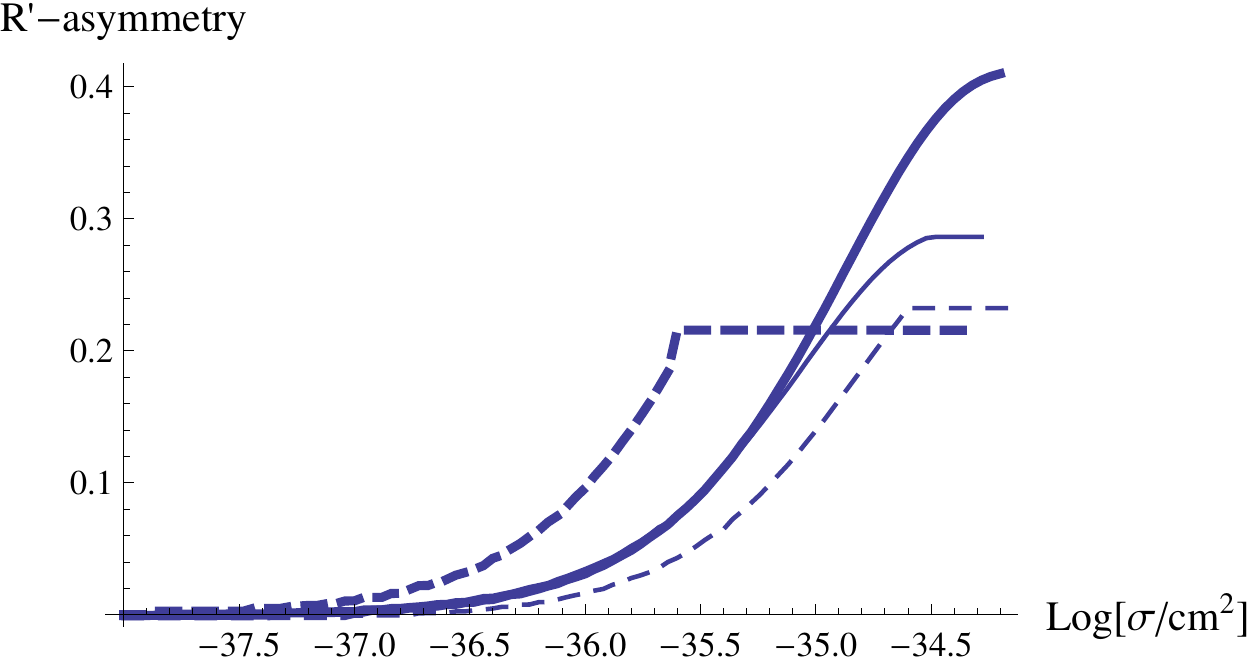}
\caption{ $R'$ asymmetry as a function of the DM-nucleon cross
section for a DM particle of mass 1 GeV at recoil energies
0.1 keV (solid thick), 0.2 keV (solid thin), and 0.3 keV (thick
dashed line). The thin dashed line corresponds to a 0.6 GeV
DM particle with recoil 0.1 keV. We assume a Na detector.} \label{fig1}
\end{center}
\end{figure}
\begin{figure}[h!]
\begin{center}
\includegraphics[width=.4
\textwidth, height=0.3 \textwidth
]{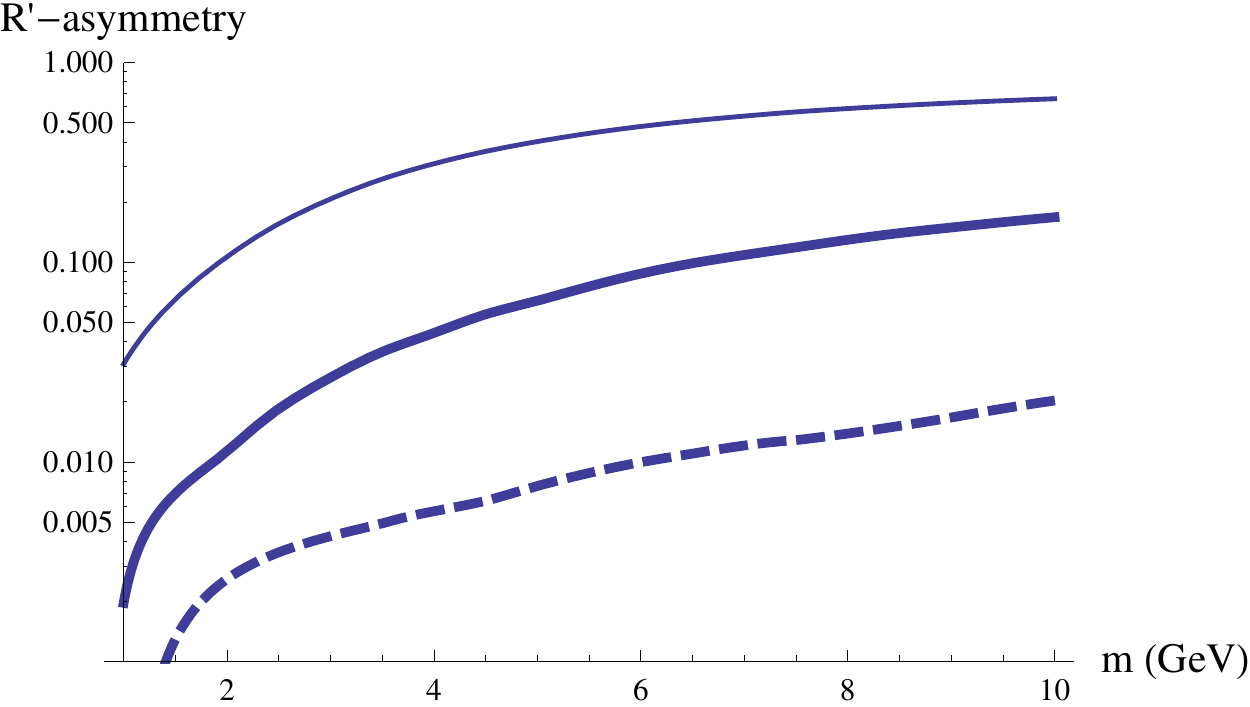}
\caption{$R'$ asymmetry as a function of DM mass (in GeV)
at a recoil energy of 0.1 keV for three values of DM-nucleon
cross section $10^{-36}\text{cm}^2$ (thin line), $10^{-37}\text{cm}^2$ (thick line) and
$10^{-38}\text{cm}^2$ (dashed line).} \label{fig2}
\end{center}
\end{figure}
\begin{figure}[h!]
\begin{center}
\includegraphics[width=.4
\textwidth, height=0.3 \textwidth
]{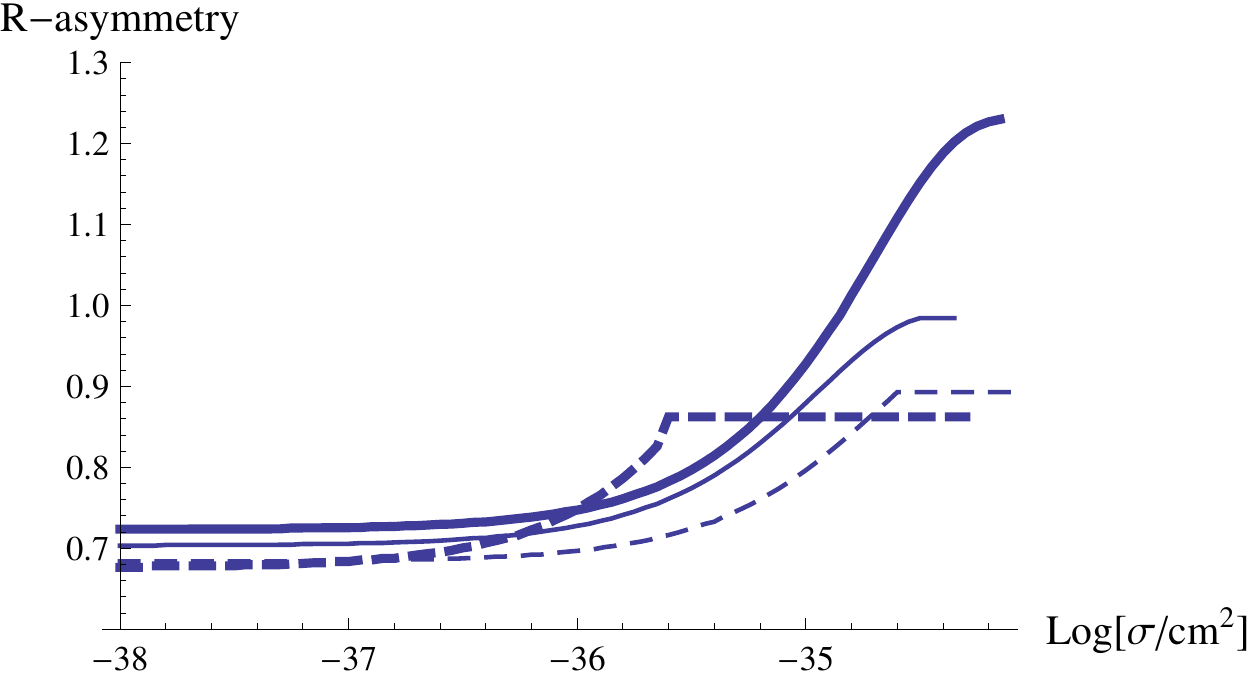}
\caption{ $R$ asymmetry for the parameters depicted in Fig.~\ref{fig1}.} \label{fig3}
\end{center}
\end{figure}
\begin{figure}[h!]
\begin{center}
\includegraphics[width=.4
\textwidth, height=0.3 \textwidth
]{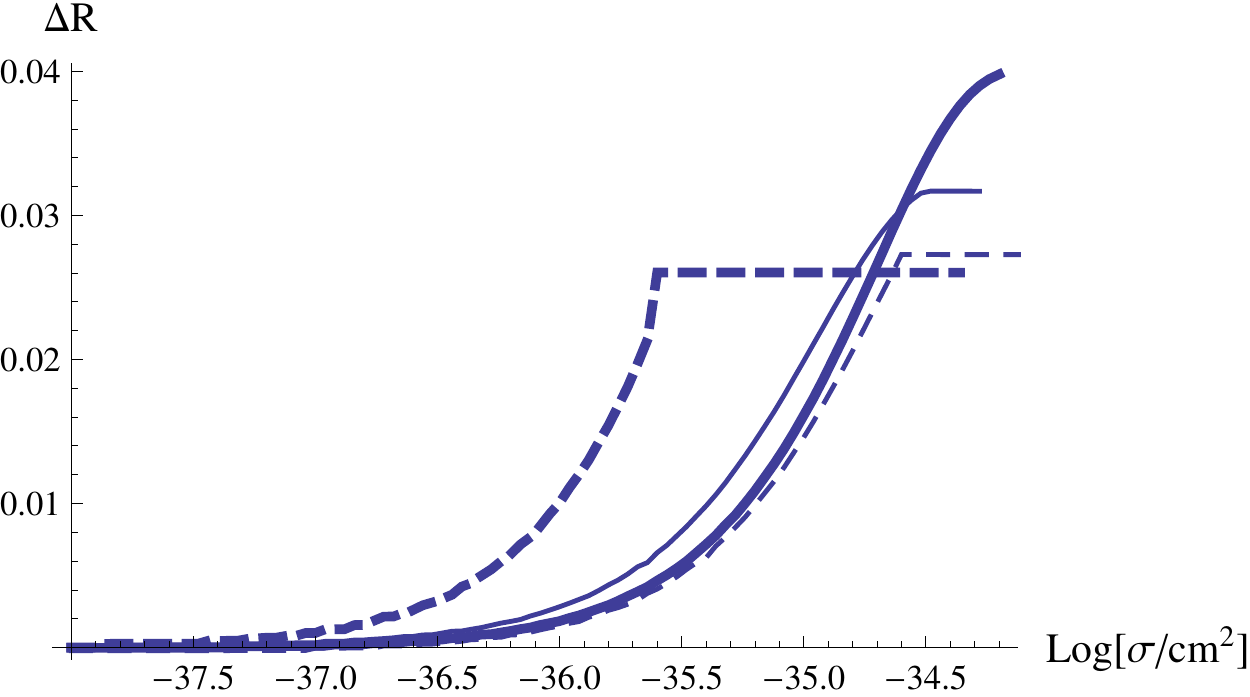}
\caption{ $\Delta R$  for the parameters depicted in Fig.~\ref{fig1}.} \label{fig4}
\end{center}
\end{figure}
\begin{figure}[h!]
\begin{center}
\includegraphics[width=.4
\textwidth, height=0.3 \textwidth
]{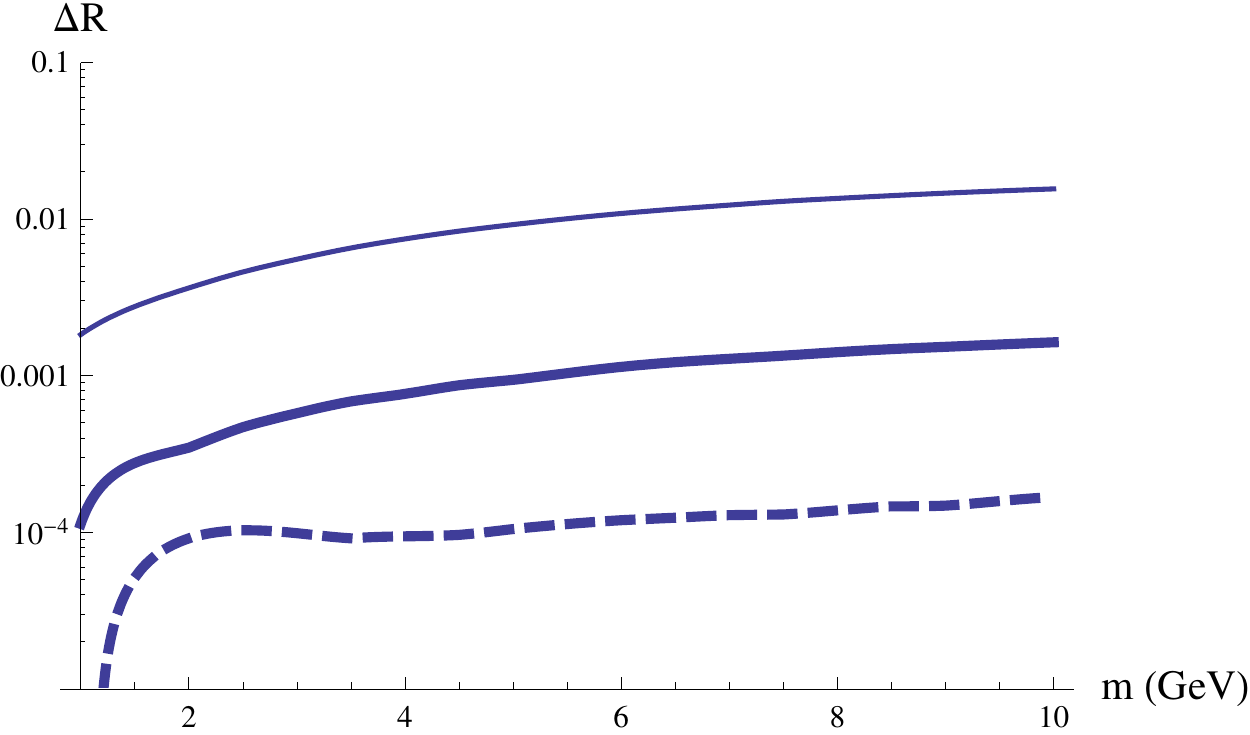}
\caption{$\Delta R$  for the parameters depicted in Fig.~\ref{fig2}.} \label{fig5}
\end{center}
\end{figure}

We present the results of $R'$, $R$ and $\Delta R$ in Figs.~\ref{fig1} to \ref{fig5}. In Fig. \ref{fig1} we show the $R'$ as a function of the DM-nucleon cross section. One can clearly see that the asymmetry increases with increasing cross section up to the point where the cross section becomes so strong that even DM particles coming from the top decelerate so much that cannot produce a recoil above the given values chosen in the figure $(\mathrm{i}.\mathrm{e}.\ 0.1,\ 0.2\ \mathrm{a}\mathrm{n}\mathrm{d}\ 0.3\ \mathrm{k}\mathrm{e}\mathrm{V})$. In addition one can notice that in the case of recoil energy 0.2 ke$\mathrm{V}$ and in a more pronounced way in 0.3 $\mathrm{k}\mathrm{e}\mathrm{V}$, the asymmetry seems to be flat for a range of DM-nucleon cross section. In reality the asymmetry drops slightly within this range until the critical value of the cross section where no particles can reach the detector from any side. The reason we have an almost flat asymmetry for that range of cross section is simple. The asymmetry increases as a function of the cross section up to the point where DM particles that come from below (traveling a distance of the earth's diameter) decelerate to low energies that cannot produce the given recoil. As the cross section increases further, the asymmetry is not affected simply because there are no more particles coming from below and therefore the asymmetry cannot increase further. The asymmetry drops only slightly because larger cross section now starts to lead to lower number of events from above. However since the distance from above is not large (we have taken a typical 1.6 km), the significant drop in the number of events happens sharply at $\sim 10^{-34}\mathrm{c}\mathrm{m}^{2}$. Fig.~\ref{fig2} depicts the $R'$ asymmetry as a function of the DM mass for three distinct values of the DM-nucleon cross section. Generally, one can conclude that especially for light enough DM particles where the allowed DM-nucleon cross section might not be so small since it is barely constrained by current direct detection experiments, the asymmetry in the up-down directional detection due to interactions of DM with underground atoms is a large fraction of the overall asymmetry that includes also the asymmetry due to the difference in the up-down DM flux caused by the motion of the earth in the galaxy. Fig. \ref{fig3} represents the same parameter space as in Fig. \ref{fig1} for the $R$ asymmetry instead of $R'$. We have chosen to show also the $R$ asymmetry because it is the ratio between the asymmetry of up-down events over the forward-backward asymmetry (which is the one between the directions $\hat{v}_{e}$ and $-\hat{v}_{e})$. This comparison is important since as mentioned earlier, it is the most thorougly studied in the case of directional detectors. Fig. \ref{fig3} also verifies the findings of the previous figures, i.e. for light DM particles with relatively strong DM-nucleon cross section, the stopping effect of underground atoms is significant. Finally Figs. \ref{fig4} and \ref{fig5} show that for DM-nucleon cross section of the order of $10^{-36}\mathrm{c}\mathrm{m}^{2}$ or larger and for DM masses of 1 Ge$\mathrm{V}$ or lower, the asymmetry can be of the order of a few percent with respect to the total non-directional detection rate. With the advent of new directional detectors with lower energy recoil thresholds, not only will be possible to probe lighter DM candidates but as we point out in this paper, we can gain significant information regarding the type and strength of DM-nucleon interactions. There is parameter phase space in the region of light DM where the stopping effect of underground atoms on DM particles might be statistically significant.

 In this paper we make a first attempt to identify the importance of the stopping effect in the context of the directional DM detectors with respect to the well studied forward-backward directional asymmetry. We assume contact type DM-nucleon interactions, and a constant density for the earth. We derive formulas that give the energy loss of DM particles as they travel underground based on coherent scattering with the oxygen nuclei abundant in earth, which are the most effective ones as long as we have spin-independent interactions. We also provide formulas that give the directional detection rate taking into account this effect assuming a typical Na detector. We propose an up-down asymmetry in the directional detection rate as the best parameter one can use to study the significance of this stopping effect. We demonstrate that this up-down asymmetry in the directional detection rate can be a few percent of the total non-directional detection rate for a large range of DM-nucleon cross section and mass, and therefore it could be observed in upcoming direct detection experiments with directional detectors. Although we have presented results for a Na detector, our results are quite generic in the sense that one can easily use our formulas for a different target nucleus. The up-down asymmetry in directional detectors has two potential sources, i.e. the stopping effect and the asymmetry in the DM flux due to the velocity of the earth with respect to the DM halo. We demonstrate that there is phase space where the stopping effect represents a significant fraction of the overall asymmetry. 

We leave several things for future work. One can in- clude other elements than just oxygen for the DM stopping effect, a non-constant density profile for the earth, and different types of DM-atom interactions. For example long range DM-atom interactions or DM-electron interactions can have a significant amount of stopping if DM particles travel through metallic layers of the earth. In principle, if sufficient number of events is detected, this technique can be used as a ``Dark matter tomography''. One could study the density and composition profile of the earth based on the directional detection rate of DM that has traveled different distances and segments of the earth's interior, given that the DM-atom interactions have been identified and understood.

The author is supported by the Danish National Research Foundation, Grant No. DNRF90. This work was partially performed at the Aspen Center for Physics, which is supported by National Science Foundation grant PHY-1066293.

\end{document}